# Unveiling low THz Dynamics of Liquid Crystals: Identification of Intermolecular Interaction among Intramolecular Modes

Patrick Friebel[a], Daria Ruth Galimberti[b], Matteo Savoini[c], Laura Cattaneo[a*]

[a] Max Planck Institute for Nuclear Physics, Saupfercheckweg 1, 69117 Heidelberg, Germany
[b] Institute for Molecules and Materials, Radboud University, Heyendaalsweg 135, 6526 AJ Nijmegen, The Netherlands
[c] Institute for Quantum Electronics, ETH Zürich, Auguste-Piccard-Hof 1, 8093 Zürich, Switzerland

[*] Email: cattaneo@mpi-hd.mpg.de

## Abstract

Liquid crystal based technologies have found considerably diversified uses and areas of application over the last few decades, proving to be excellent materials for tunable optical elements from visible to near-infrared frequencies. Currently, much effort is devoted to demonstrating their applicability in the far-infrared or THz spectral frequency (1 - 10 THz), where tremendous advances have been achieved in terms of broadband and intense sources. Yet a detailed understanding of the dynamics triggered by THz light in liquid crystals is far from complete. In this work, we perform broadband THz Time Domain Spectroscopy on the low-frequency modes of 4-cyano-4′-alkylbiphenyl (nCB) and 5-phenylcyclohexanes (PCH5) across different mesophases. DFT calculations on isolated molecules capture the majority of the measured response above 3 THz. In particular, the pronounced modes around 4.5 and 5.5 THz mainly originate from in-plane and out-of-plane bending of the cyano group. In contrast, the broad response below 3 THz, linked to modes of the alkyl chain, disagrees with the single molecule calculation. Here, we identify a clear intermolecular character of the response, supported by dimer and trimer calculations.

# Introduction

Liquid Crystals (LCs), prominent examples of soft matter materials, exhibit mesophase character, combining different solid and liquid properties within the same phase. Since their discovery in 1888 by Friedrich Reinitzer,[1] LCs found ever-increasing interest from science and technology thanks to their unique properties in terms of birefringence, polarizability, and their response to external stimuli, like electric and magnetic fields. Most commonly known are the calamitic thermotropic nematic and smectic A phases (see Figure 1b). Nematics exhibit orientational order, meaning their long axis is oriented along a collective director $\vec{n}$ within a macroscopic volume, while retaining all degrees of freedom in terms of motion, i.e., they flow like a liquid. Smectics (A denoting the subcategory with the lowest degree of order) form layers in which the molecules point along the layering direction and can freely move within each layer, i.e., they exhibit orientational order, and positional order along the pointing direction. Given the liquid nature of LC materials, these degrees of order are far from perfect. To quantify this, a scalar parameter, called orientational order parameter ($S$), is often introduced:[2] $S=\langle 3(\cos\theta)^2-1\rangle/2$, where $\theta$ denotes the polar angle of each molecule to the director, and the brackets express the volume averaging. $S = 0$ describes total disorder (isotropic state) and $S = 1$ describes total order (perfect crystal). For nematic and smectic phases $S \approx 0.3 - 0.8$ are typical values.[2]

Past and current LC-based applications rely on the stabilization of a desired order parameter within large temperature ranges, and on the alignment of the LC director via moderate electric fields, thus the optical axis along a specific direction. In fact, LCs developed into ideal candidates for optical devices, such as tunable wave plates[3,4] and phase masks[5] in the visible to infrared (IR) wavelengths. The steady advancements in tabletop THz sources led to intense ongoing research on possible devices and materials to modulate and control the THz waves in a similar manner. Nematic LCs were already demonstrated to be low-loss and tunable birefringent materials for THz-wave plates and phase shifters.[6-9] However, their applicability is still limited because of the counteracting combination of birefringence and necessary thickness of the medium.[10,11] The wavelengths associated with THz radiation (0.1 - 10 THz) are long (3000 - 30 μm). To obtain a reasonable phase shift $\delta$, the LC thickness has to become considerably larger in the THz range compared to optical frequencies, and losses become no longer negligible. The ideal LC presents high birefringence and full transparency (low losses) at THz frequencies. To design new switching devices or guide molecular design targeting such THz-performances, we need a detailed understanding of the microscopic properties of the LC in the THz range.

Vieweg et al.[12,13] reported one of the first studies on frequency dependent refractive indices and absorption coefficients in the nematic liquid crystals of the nCB family (4-cyano-4′-alkylbiphenyl) in the frequency range from 0.3 to 15 THz using THz Time Domain Spectroscopy (TDS) in the nematic phase. The observed spectra are dominated by multiple spectral features mainly at frequencies above 4 THz, assigned to either intra- or intermolecular vibrations and Poley-type absorption, i.e. the torsional oscillation about the rigid long axis of the LC molecule itself.[14,15] However up to now, a full understanding of these signatures is still to be achieved.

Beyond the technological target, such understanding is deeply connected to more fundamental questions: How does the molecular structure of the LC compound affect the light-matter interaction in the low THz range? What is the impact of the LC phase, thus the collective intermolecular environment, on the dynamical properties in this low-frequency regime? In this work, we investigate the absorption features in the 1 to 7.5 THz frequency range of different 4-cyano-4′-alkylbiphenyl (nCB) and 5-phenylcyclohexanes (5PCH) at different temperatures, i.e., mesophases, by exploiting their structural similarities in order to gain insight into the modes associated with each molecular building block. Gas phase DFT calculations have been performed to unfold the features and provide a detailed interpretation of the observed spectra.

## Methods

*LC materials* We choose 8CB (4-Octyl-4'-cyanobiphenyl) and other representatives (see Table 1, Synthon Chemicals, LC used as received without further purification) of the nCB series (n indicates the number of carbon atoms in the alkyl chain) as our primary samples for multiple reasons. Firstly, they are one of the most widely studied and used LC in science and technology, mostly driven by the fact that they are commercially available and show mesophases close to room temperature. Secondly, their structural similarity to PCHn (trans-4-(4′-n-alkylcyclohexyl)benzonitrile) allows us to make direct comparisons and disentangle structural contributions in the measured response.

*Table 1: Liquid Crystal compounds under investigation. All LCs are commercially available from Synthon Chemicals. The phase transitions and purities are taken from the product information. Phases: Cr - crystalline, SmA - smectic A, N - nematic, I - isotropic*

| Sample | Full Name | Phase Transitions | Purity [%] |
|---|---|---|---|
| 4CB | 4-Butyl-4'-cyanobiphenyl | Cr - 48°C - I | 99.5 |
| 5CB | 4-Pentyl-4'-cyanobiphenyl | Cr - 24°C - N - 35.5 - I | 99.5 |
| 8CB | 4-Octyl-4'-cyanobiphenyl | Cr - 21.5°C - SmA - 33°C - N - 40.5°C - I | 99.5 |
| 10CB | 4-Decyll-4'-cyanobiphenyl | Cr - 44°C - SmA - 51.1 - I | 99 |
| PCH5 | trans-4-(4´-n-Pentylcyclohexyl)benzonitrile | Cr - 31°C - N - 55°C - I | 99.5 |

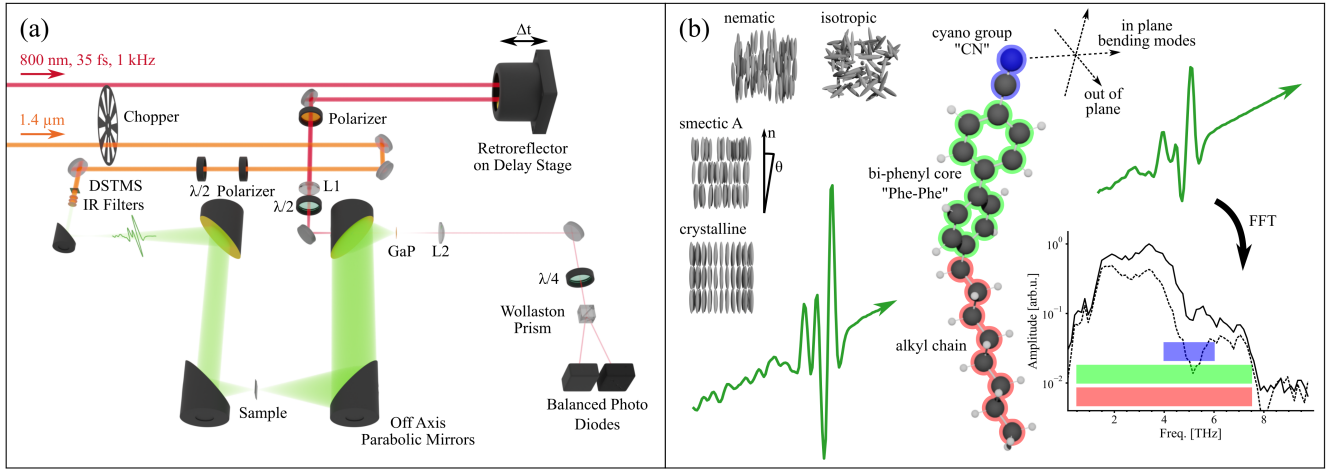

*Figure 1: (a) Sketch of the Time Domain Spectroscopy setup. Broadband THz pulses are generated by optical rectification of 1.4μm radiation in DSTMS. The THz is guided into the sample using parabolic mirrors in nitrogen purged atmosphere. The transmitted pulses are then focused in 200 μm thick GaP for Electro Optic Sampling, measuring copropagating 800 nm, 35 fs light in balanced detection as a function of time delay. (b) Top left: sketch of all four phases under investigation with rod-like molecules oriented along a common director n and random fluctuations θ. Centre: pictorial view of the experiment. The generated broadband pulses are propagating through the molecule (here 8CB) along their ordinary or extraordinary axis, and the transmitted pulses, measured in time domain, encode the sample absorption in the modulation of the pulse shape. The different structural components contributing to the response and their in-plane/out-of-plane modes, assigned by calculations, are highlighted. Bottom right: the generated and transmitted THz spectra. The colored bands indicate which molecular structures are active.*

*LC cells* The samples are prepared in sealed cells using THz transparent window material of 1mm thickness (COC polymer windows, trade name "TOPAS", microfluidic ChipShop GmbH). No surface treatments were performed on the substrates, except standard cleaning in KOH solution (1mol) bath, followed by sonication in milli-Q water (x3), acetone, and isopropanol. The sample thickness is defined by wire spacers (0.5 mm) that also act as electrodes, at which an AC electric field (1 kHz sine wave, 1 V/μm) is applied across the sample over 2mm distance in order to orient and align the sample parallel to the window surfaces. The ordinary and extraordinary axis are exposed to the incoming THz radiation by turning the cell with respect to the THz polarization, while always in good thermal contact with the temperature-controlled stage (Linkam LTS120E, temperature stability of 0.1 °C). When ramping the temperature, the rate was set to 1 °C/min and the sample was given sufficient amount of time (~ 5 min) to thermalize.

*Time Domain Spectroscopy (TDS)* We perform standard TDS on our LC samples.[12,16] The experimental setup is depicted in Figure 1a. The few-cycle THz pulses are generated by optical rectification of 1.4 μm pulses in the organic crystal DSTMS (6 mm, Rainbow Photonics). The IR pump of maximally 1 mJ is generated in an OPA (TOPAS, Light Conversion) with 800 nm NIR pulses of 35 fs duration originating from a Ti:Sa Laser System (AstrellaHE, Coherent). Using 0.5 mJ/cm$^2$ IR fluence into the DSTMS, the generated broadband THz spectrum spans 0.1 - 7.5 THz, and is therefore able to fully capture the low frequency modes in the LCs under investigation (see Figure 1b). The THz path, comprising 5 parabolic mirrors, contains the sample in the second focal plane of the set up, passing through the aperture of the temperature controlled stage, after which the transmitted THz pulse is focused into a 200 μm GaP crystal. To perform Electro-Optic Sampling, copropagating 35 fs NIR pulses taken before the OPA are measured in a balanced detection scheme as a function of time delay to the THz pulses. The volume in which the THz radiation propagates is purged with nitrogen to below 3% relative humidity in order to avoid absorption by water vapor.

*Computational details* All the spectra discussed in the paper have been computed with the Gaussian16 code.[17] A conformational search was run using the Gaussview software[18] and the MMFF4 force field as a first step for all the analyzed cases, imposing an energy window of 3.5 kcal/mol. From the generated pool of structures, in the case of the single molecule calculations, all the conformations have been retained for the next step, while in the case of the 5CB dimer and trimer, due to the computational cost, the structures with an energy of more than 2 kcal/mol from the global minimum have been discarded.

We then proceeded to a geometry optimization at the DFT level for all the selected conformations. In the case of the single molecules, the B3LYP functional,[19,20] augmented with the Grimme D3 dispersion term,[21] and the 6-311++G** basis set have been chosen. Because our calculations on the single molecule demonstrate that the BLYP[19,22] (still augmented with the D3 correction) functional has almost the same quality as the more computationally expensive B3LYP-D3 ones, in the case of the 5CB cluster calculations (dimers and trimers), the former has been preferred as a compromise between accuracy and computational cost.

As a last step, we computed the spectra for all the optimized structures in double harmonic approximation at the same level of theory used for the geometry optimization. The final spectra shown in the paper have been obtained as a weighted average by the Boltzmann populations (at 323 K) of the set of computed spectra.

The ordinary and extraordinary components of the IR adsorption spectra have been computed projecting the induced dipole moment respectively on the direction perpendicular and parallel to the static dipole moment of the molecule.

## Results and Discussion

Figure 2 shows the absorption spectra measured by TDS for 8CB with the THz polarization oriented along the ordinary (a) and extraordinary (b) LC axis. The traces are measured for different temperatures (colored lines) spanning all condensed phases presented by the compounds. Additionally, we plot the computed spectra (black line) for comparison. The computed frequencies have been scaled by a factor 1.075, and the intensities normalized on the strongest mode (5.5 THz of 8CB) to help the comparison with the experiment. 8CB is of particular interest since it is the shortest nCB showing both smectic A (21.5 - 33 °C, green solid line) and nematic (33 - 40.5 °C, orange solid line) phases. Hence, it allows us to measure any possible impact of the molecular ordering on the spectral absorbance. When looking at the spectra as a function of temperature we can see that the crystalline, smectic, and nematic phases, i.e., where the molecules are aligned, the TDS spectra are nearly identical. A more substantial difference can be spotted in the isotropic phase (red solid line), i.e., when the molecules are not aligned.

Looking at the experimental spectra obtained in 8CB with THz polarization along the ordinary axis shown in Figure 2a we can recognize four regions that we have assigned to specific modes thanks to the computational support. Below 3.5 THz the 8CB molecules show mainly (NCPhe)-Phe-alkyl skeleton modes, i.e., the CN group is not vibrating (see Figure 1b).

To support this statement, we computationally increase the atom mass (100 amu) of selected groups of the 8CB molecule and look at the impact on the absorption spectrum, as depicted in Figure 2c. When the atoms with increased mass are contained to the CN group, the corresponding absorption spectrum below 3.5 THz is almost identical to the unmodified 8CB response (see solid orange line compared to black). The small shift and change of intensities are due to the small variation of the modes reduced mass. On the other hand, the spectrum completely lacks all modes above 3.5 THz, where the CN group plays a dominant role. Further increasing the number of atoms with heavier mass, like in our case CCN (light purple line) and C-(CCN)-C (dark purple line), affects the entire spectrum, including below 3.5 THz, showing the delocalized and collective nature of this mode.

Between 4 - 6 THz, the CN group takes part in the vibration. The alkyl chain skeleton modes are now coupled with the CCN *bending* and the *biphenyl (Phe-Phe) bending* modes. The out-of-plane bending modes tendentially are located at lower frequencies (4-5THz) compared to the in-plane bending modes (5-6 THz) as indicated by the computed polarized spectra in Figure 2d on the direction parallel (orange curve) and perpendicular (purple curve) to the phenyl rings. Finally, between 6 - 8 THz mainly (NCPhe)-Phe-alkyl skeleton modes (no vibrations of the CN group) can be found.

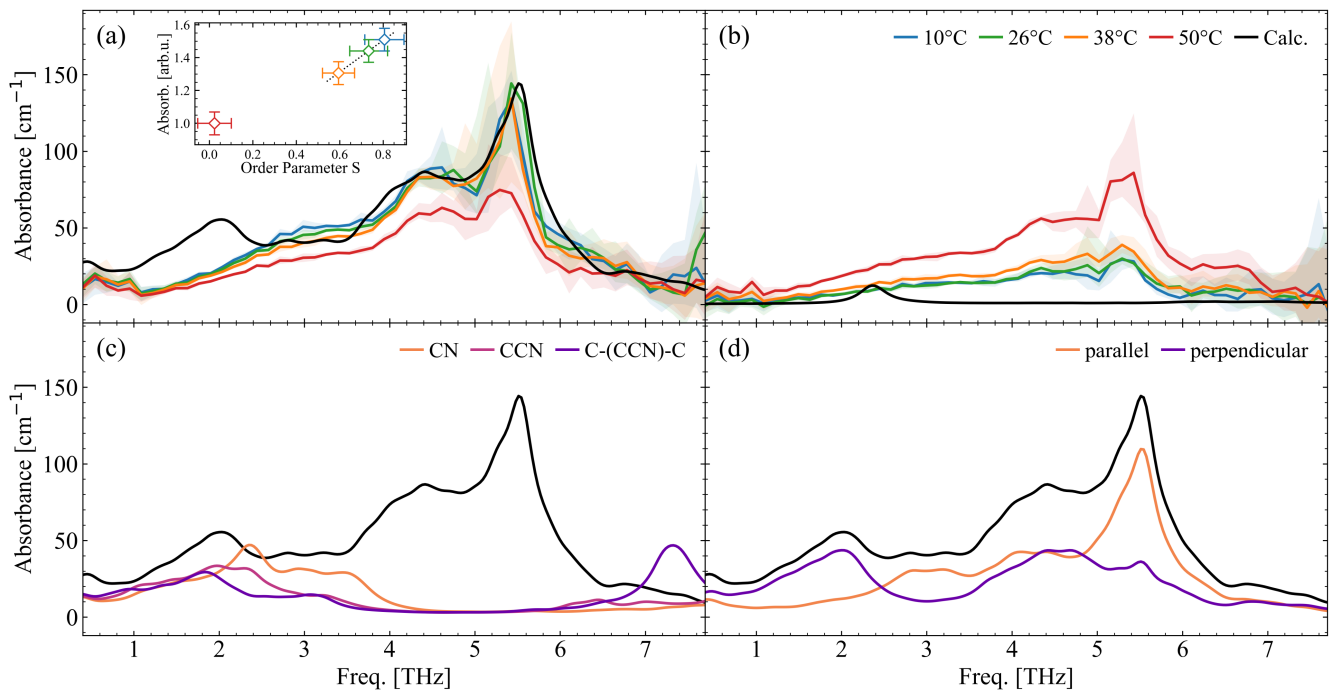

*Figure 2: Absorption spectra of 8CB obtained by TDS along (a) the ordinary and (b) the extraordinary axis. The spectra (colored traces) have been measured for sample temperatures of 10 °C, 26 °C, 38 °C, and 50 °C, corresponding to crystalline, smectic A, nematic, and isotropic phases. Inset in (a): Estimation of the degree of order S compared to the intensity of the lowest absorption band. Black line: Harmonic gas-phase computed spectra. To help the comparison with the experimental data, the computed frequencies have been scaled by a factor of 1.075 and the intensities normalized on the 5.5 THz peak. (c) Computational deuteration of isolated components of the molecule and the resulting response when increasing the atomic mass (100 amu), effectively killing their contributions. When the heavy atoms are contained to the CN group (orange) the remaining response resembles the low frequency modes. Further increasing the heavy region to CCN (light purple) and C-(CCN)-C (dark purple) starts to modify also those modes. (d) Calculated ordinary response (black) and its contributions when polarized parallel (orange) and perpendicular (purple) to the phenyl rings.*

The comparison between the experimental and the calculated single molecule spectra shows very good agreement above 3.5 THz. At the same time, the crystalline, smectic, and nematic phase experimental spectra above 3.5 THz show only negligible differences (within the experimental uncertainty), and the differences with the isotropic phase originate from the alignment. Therefore the absorption can be attributed to the ensemble of individual vibrating molecules, in other words, mainly an intramolecular response.

The situation changes at and below 3.5 THz. We observe a clear reduction in total intensity with increasing temperature. We can see that by plotting the integrated intensity of the lowest ordinary mode as a function of the experimentally-derived order parameter S (see inset in Figure 2a) we find a linear trend (excluding the isotropic phase). At the same time, discrepancies between measurement and gas phase calculation are evident. This points towards an effect of the intermolecular order. Interestingly, the peak we find at 2 THz in the computed spectra is almost completely explained by modes pointing out of the plane of the ring (purple curve in Figure 2d). Given the preferential co-facial arrangement of neighboring molecules, in which the bi-phenyl structures face each other in an antiparallel manner[23,24] to maximize both the dipole-dipole interactions and the π-π stacking, it is reasonable that these modes are most sensitive to intermolecular interactions and that they are modulated by this kind of packing in a condensed phase. We qualitative test this assumption using dimer and trimer calculations at the end of this section.

Looking now at the 8CB extraordinary spectra, we can first observe how it decreases with lowering temperature, whereas the total intensity between 2 - 8 THz increases in the ordinary absorbance. Interestingly the calculations barely predict any absorption, while the measurements show a clear response. In particular, below 3.5 THz the calculations predict only one vibrational mode generating dipole fluctuations parallel to the molecular axis, i.e., active along the extraordinary axis. This mismatch between experiments and calculations is explained by considering the molecules' non-perfect alignment in the LC cell, even in the presence of an aligning AC field. Given the typical order parameter $S$ of LC below one (for nematic and smectic phases $S \approx$ 0.3 - 0.8), as previously introduced,[2] we are most likely probing projections of the ordinary axis of molecules slightly misaligned with the director, giving contributions to the extraordinary absorbance. Coherent with this, we see little difference in the lowest temperature extraordinary traces, associated with the crystalline and smectic A phases (blue and green solid line in Figure 2b), and then an increase for the nematic case (orange solid line). It has been shown in the past that the smectic phase forms ordered bilayer structures.[24] Inside the crystalline and smectic A phases ordered packing, where the bi-phenyl structures are stacked next to each other, the rotational modes are more hindered compared to the nematic phase, where there is also the freedom for the molecules to be randomly translated with respect to each other along their long axis.

In order to further confirm the predicted contributions of major structural components of the molecule to the response in the different spectral ranges, and to identify the reason for the discrepancy at lower frequencies, we follow two strategies. Firstly, we compare measurements of further members of the nCB and PCHn families of LC across different mesophases and molecular orientations, giving us the ability to distinguish contributions of the structure by changing the tail length and the core composition, respectively. Another approach is to computationally extend the calculations by building clusters of two (dimer) or three (trimer) molecules. The idea is to find indications of intermolecular interactions modifying the absorption; hence in a first-order manner, we are taking into account the collective environment of our samples.

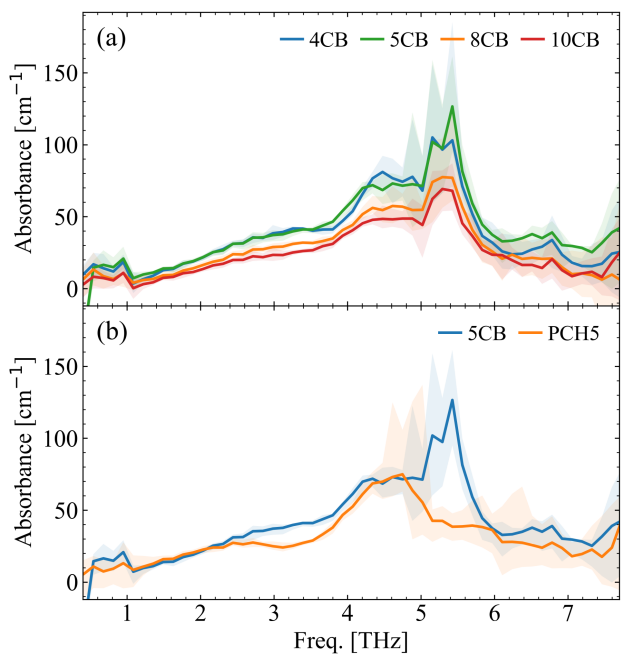

*Figure 3: (a) Absorption spectra of selected members of the nCB series in isotropic phase measured by TDS. Since the chosen molecules show different phase behavior, the isotropic phase is the only one they have in common. The temperatures are set to 10°C - 20°C above the phase transition. The comparison allows for investigation of the effect of the alkyl chain length. (b) Absorption spectra of 5CB and PCH5 in the isotropic phase. The comparison allows to identify the contribution of the core towards the molecules THz activity while the alkyl chain length is held constant.*

In Figure 3a we consider the isotropic phase, where the system is in a disordered state and behaves liquid-like for the nCB family. Given the different phase behavior among the nCBs, the isotropic phase is the state all have in common, allowing for comparison across the widest range in the chain length. The first thing that we notice is that by increasing the chain length, we see a steady decrease in absorbance for the whole 1 - 7.5 THz region. This is not surprising considering that the intensity is measured per unit volume, and a decrease in the number of molecules per unit volume is expected with increasing chain length due to the pure steric effect.[25,26] However, the trend is not equal for all the spectral features. To explain this, the first factor we should consider is that the modes have different origins.

Increasing the chain length, the number of CN per unit volume decreases, bringing a decrease in intensity for those spectral features dominated by the cyano group, thus between 4 and 6 THz. At the same time, the number of $CH_2$ groups per unit volume increases with the chain length, partially counteracting the decreased density, especially below 3.5 THz, where skeleton modes of the alkyl chains are massively involved. Therefore, modification of the chain length mainly serves as a tunable parameter affecting the overall amplitude of the response in the region under investigation. The density of the material and the number of $CH_2$ compete to determine the final integrated intensity.

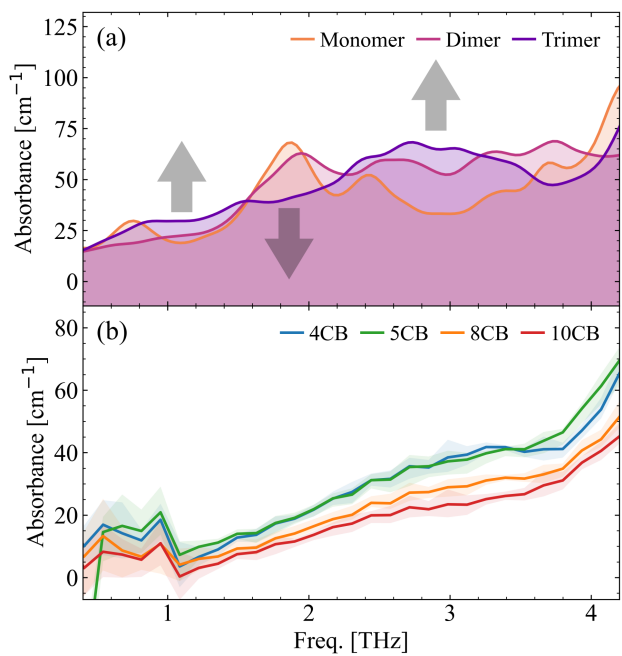

*Figure 4: (a) Calculated spectra for monomer (orange), dimer (light purple), and trimer (dark purple) of 5CB at BLYP level. A clear shift of the pronounced low frequency peak towards higher frequency and broader width with increasing molecule number is visible, indicating an intermolecular response. (b) The low frequency response of selected members of the nCB series at isotropic mesophase for comparison with (a).*

In the next step, we compare 5CB and PCH5 in their isotropic phases, yielding information on the effect on the absorbance when one phenyl ring is exchanged with a cyclohexane group as shown in Figure 3b. While in the case of the nCB series, the two rings almost lie in the same plane, in the case of PCH5 they lie at 90° with respect to each other. We observe clear modification of the absorption behavior in this region. The cyclohexane ring is much less structurally rigid compared to a phenyl ring, leading to more degrees of freedom for the molecular skeleton to move. As a consequence, we see a strong suppression/modification of the 5.5 THz mode and the creation of a valley at 3.2 THz within the broad low-frequency band. The mode at 4.5 THz is instead largely unaffected within the measurement uncertainty, with a small narrowing as the only noticeable change. The calculations show that this is due to the balance between two opposite effects: (i) The missing coupling between the two rings bending modes red shift the in-plane bending to 4 - 5 THz, i.e., the in-plane and out-of-plane modes are almost degenerate. (ii) The absolute intensity of each mode on average decreases, possibly, because the missing coupling between the two rings reduces the charge fluxes responsible for the IR intensity. We can conclude that the vibrational response of the material can be indeed tuned by the modification of the polar head.

Lastly, we discuss the discrepancy between measurement and calculation below 3.5 THz by comparing calculations with an increasing number of molecules. Past work ascribed the low THz mode to Poley absorption,[15,27,28] and later to the Ioffe-Regel crossover[13] in relation to disordered systems. Here we show that a much simpler explanation, based on conventional intermolecular interactions, should be considered. In Figure 4a we use 5CB as a test molecule because it comes with reduced computational cost compared to 8CB, meaning only 27 possible single conformations in 5CB compared to 264 in 3.5 kcal/mol for 8CB. The computed data show that the intermolecular interactions are responsible for the blue-shift of the 2 THz modes in the clusters (dimers and trimers) with respect to the gas phase. These blue shifted modes compare favorably on qualitative terms with the measured responses shown in Figure 4b. While this model is a simplification of the molecular packing in the condensed mesophases, it gives insight into the intermolecular character of the mode on display.

## Conclusion

We performed THz TDS measurements on 8CB and compared them with single molecule DFT calculations. While the calculation well describes the majority of the LC response, thus it behaves as an ensemble of single molecules, in the low frequency range of the spectrum (< 3.5 THz) we find significant disagreement with this picture. From the analysis of the modes parallel and perpendicular to the phenyl rings in the ordinary response, and by computing dimer and trimer configurations, we find clear markers of the intermolecular nature of this mode. By comparing the 8CB measurements to other molecules in the nCB series, and by comparing 5CB to PCH5, we are able to disentangle the contributions of the alkyl chain and of the core to the THz response. The chain has only an effect on the global intensity of the spectrum, while changes to the core can lead to significant shifts of certain modes.

Our results indicate that while for traditional applications, such as wave retarders in the THz regime, the absorbing character of the LCs under investigation pose a major challenge, novel approaches to control the material properties in THz fields might be open. While the absorption spectrum in this regime does not show any significant change of features as a function of the LC phase, we are able to distinguish molecular responses dominated by intramolecular vibrations from modes influenced by intermolecular interactions in condensed phase. Given the separation of these modes, it's possible to selectively drive a given molecular response by the right choice of THz field. Alternatively, by the choice of the molecular structure, such as the less rigid core of PCHn compared to nCB, it is possible to shift the dominant mode of the response. This enables us to modify material properties, such as transiently induced modulation of birefringence on ps time scales,[29] in a systematic manner, paving the way for new tools in ultrafast light control.

## Author Information

### Corresponding Author

Laura Cattaneo,

Max Planck Institute for Nuclear Physics, Saupfercheckweg 1, 69117 Heidelberg, Germany,

Email: cattaneo@mpi-hd.mpg.de

ORCID: https://orcid.org/0000-0001-7492-3850

## Notes

The authors declare no competing financial interests.

## Acknowledgements

L.C. would like to acknowledge the Max Planck Group Leader program for funding her independent research. D.G. thanks SURF (www.surf.nl) for providing computing time and for the support in using the Dutch National Supercomputer Snellius.